\documentclass[aps,pre,onecolumn,superscriptaddress,notitlepage]{revtex4-1}

\usepackage[a4paper, margin=2.50cm]{geometry}

\usepackage{amsmath}
\usepackage{graphicx}
\usepackage{fancyhdr, color}
\usepackage{bm}
\usepackage{accents}
\usepackage{amssymb}
\usepackage{url}
\usepackage{setspace}
\usepackage[dvipdfmx]{hyperref}

\newcommand{\beq}{\begin{equation}}
\newcommand{\eeq}{\end{equation}}
\newcommand{\beqa}{\begin{eqnarray}}
\newcommand{\eeqa}{\end{eqnarray}}
\newcommand{\Tr}{\text{Tr}}

\newcommand{\av}[1]{\left\langle #1 \right\rangle}

\newcommand{\aav}[1]{ \hspace{-.6mm}\left<\hspace{-.4mm} \right. #1 \hspace{-.9mm}\left. \right> }

\newcommand{\ee}{\mathrm{e}}

\begin{document}

\fancyhead[L]{}
\fancyhead[C]{ \color[rgb]{0.4,0.2,0.9}{Quantum Thermodynamics book}}
\fancyhead[R]{}

\title{Quantum Fluctuation Theorems}

\author{Ken Funo}
\email{kenfuno@pku.edu.cn} 
\affiliation{School of Physics, Peking University, Beijing 100871, China}

\author{Masahito Ueda}
\email{ueda@cat.phys.s.u-tokyo.ac.jp}
\affiliation{Department of Physics, The University of Tokyo, 7-3-1 Hongo, Bunkyo-ku, Tokyo 113-0033, Japan}
\affiliation{RIKEN Center for Emergent Matter Science (CEMS), 2-1 Hirosawa, Wako, Saitama 351-0198, Japan}

\author{Takahiro Sagawa}
\email{sagawa@ap.t.u-tokyo.ac.jp}
\affiliation{Department of Applied Physics, The University of Tokyo, 7-3-1 Hongo, Bunkyo-ku, Tokyo 113-8656, Japan}

\date{\today}

\begin{abstract}
Recent advances in experimental techniques allow one to measure and control systems at the level of single molecules and atoms. Here gaining information about fluctuating thermodynamic quantities is crucial for understanding nonequilibrium thermodynamic behavior of small systems. 
To achieve this aim, stochastic thermodynamics offers a theoretical framework, and nonequilibrium equalities such as Jarzynski equality and fluctuation theorems provide key information about the fluctuating thermodynamic quantities. We review the recent progress in quantum fluctuation theorems, including the studies of Maxwell's demon which plays a crucial role in connecting thermodynamics with information.
\end{abstract}

\maketitle

\thispagestyle{fancy}


\section{Introduction}

The fluctuation theorem (FT) may be regarded as a modern clue to the problem raised by Loschmidt, who posed a serious question about irreversible processes in time-reversal-symmetric dynamics~\cite{Loschmidt}. In accordance with time-reversal symmetry, the entropy production can be negative albeit an exponentially small probability~\cite{Seifert12,Jarzynskireview}, which provides a new insight into our understanding about the arrow of time~\cite{Jarzynskireview,Campisi11arrow}. The averaged entropy production is thus always nonnegative, consistent with the second law of thermodynamics. 

The FT reveals fundamental properties of the entropy production under nonequilibrium dynamics, which has opened up the field of stochastic thermodynamics~\cite{Seifert08review,Sekimoto,Seifert12,Jarzynskireview}. Various types of FT have been discussed in literature~\cite{Evans,Gallavotti,Jarzynski1,Crooks,Jarzynski00,HatanoSasa,Seifert05,Esposito10,Kawai}, and they can be obtained in a unified way by starting from the detailed FT~\cite{Seifert12}. In particular, the Jarzynski equality~\cite{Jarzynski1} and the Crooks FT~\cite{Crooks} allow one to determine the equilibrium free energy through measurements of nonequilibrium work~\cite{Jarex,Fluctex}. Experimentally, the classical FT is relevant to classical small systems such as biomolecules, molecular motors, and colloidal particles, while its quantum counterpart is relevant to quantum devices such as NMR systems~\cite{Batalhao14}, trapped ions~\cite{An15}, and superconducting qubits~\cite{Naghiloo17}. 


In this article, we first review the quantum FT by focusing on the Jarzynski equality and the Crooks FT~\cite{Esposito,fluctuation1}. These relations are applicable to externally driven quantum systems far from equilibrium, and are thus relevant to quantum devices with rapid external control. 
The quantum FT has been formulated for isolated systems described by unitary dynamics~\cite{Kurchan,Tasaki,Talkner07,Deffner11FT}. Further studies have been carried out, including the quantum FT for open quantum systems~\cite{Crooks08,Monnai,Jarzynski04b,Campisi09,Horowitz12,Hekking13}, monitored quantum systems~\cite{Cyril17,Lutz16,Campisi10}, and quantum field theories~\cite{Caselle16,Caselle18,Deffner18}. In particular, the quantum jump method allows one to assign quantum work and heat along individual quantum trajectories, as in the case for classical Markov jump processes~\cite{Horowitz12,Hekking13}. For simplicity, in this article we only consider the case with a single heat bath, though the extension to the case with multiple heat baths is straightforward. This setup includes applications to quantum heat engines and quantum heat transports~\cite{Gaspard,Saito,Saito2,Nakamura}.  


If we can access thermal fluctuations of the system via measurement and feedback control, we can demonstrate the fundamental connections between the thermodynamic properties and the information-theoretic quantities. This setup is a modern formulation of Maxwell's demon~\cite{Maxwell,Maruyama}, opening an interdisciplinary field of  information thermodynamics~\cite{Sagawa12review,Sagawa12reviewa,Parrondo}. 
The fundamental bound on the capability of Maxwell's demon has been revealed in the form of the generalized second law by including the information content~\cite{Sagawa08,Sagawa09}. A generalized FT under measurement and feedback control has been derived in both classical~\cite{Sagawa10,Sagawa12} and quantum~\cite{Morikuni,Funo13,Funo15} regimes.


This article is organized as follows. In Sec.~\ref{sec:FTW}, we review the second law and the FT in the quantum regime, for both isolated and open systems. In Sec.~\ref{sec:FTF}, we review the case of measurement and feedback control by Maxwell's demon. In Sec.~\ref{sec:experiment}, we comment on related experimental studies. In Sec.~\ref{sec:summary}, we make concluding remarks.

\section{\label{sec:FTW}Second law and fluctuation theorems}

In this section, we utilize techniques in quantum information theory to derive the second law of thermodynamics, which sets a fundamental bound on the entropy production. We then discuss a stochastic version of the entropy production along individual quantum trajectories, and use it to derive the quantum FT.  In Sec.~\ref{sec:a} and Sec.~\ref{sec:FT}, we consider a unitary time evolution of the composite system of a driven system and the heat bath. In Sec.~\ref{sec:c}, we consider Gibbs preserving maps and open quantum systems and discuss the quantum jump method to derive the quantum FT. See also Sec.~III and Sec.~V of Ref.~\cite{QbookCH27} for the derivations of the second law in setups similar to those of Sec.~\ref{sec:a} and Sec.~\ref{sec:c} of this chapter.

\subsection{\label{sec:a}Derivation of the second law of thermodynamics}

\subsubsection{\label{sec:setup}Setup}

\begin{figure}[t]
\begin{center}
\includegraphics[width=.6\textwidth]{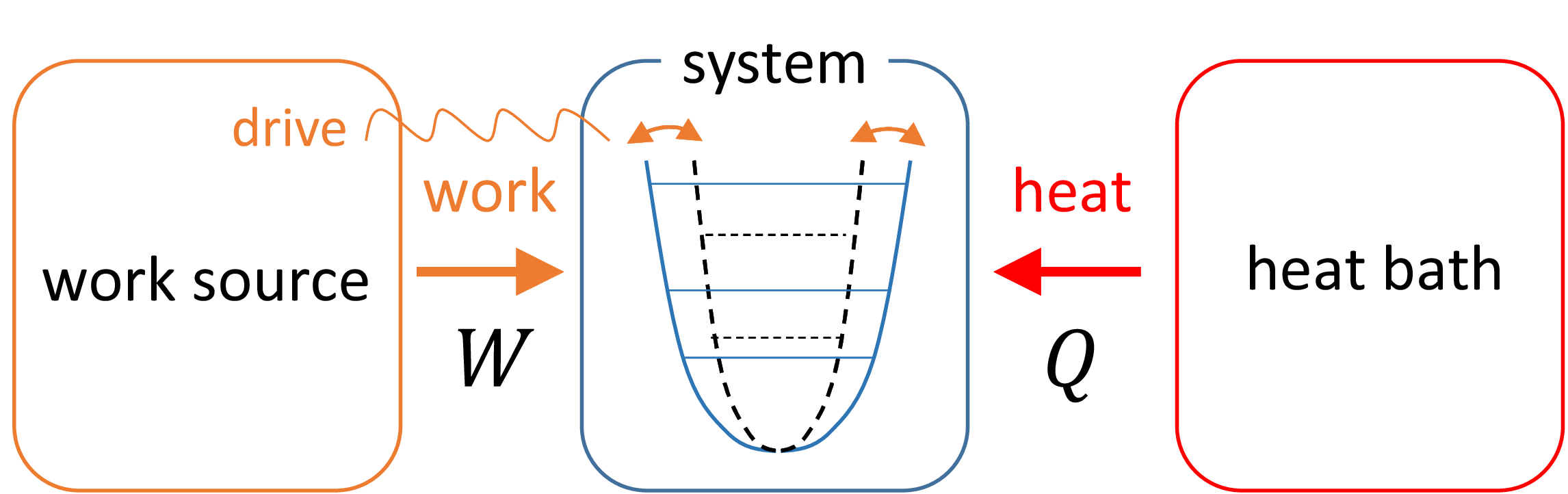}
\caption{ Setup of the quantum fluctuation theorem. 
The Hamiltonian of the system is modulated by an external drive, which performs work $W$ on the system. In this process, the system absorbs heat $Q$ from the heat bath.
}
\label{fig:setup}
\end{center}
\end{figure}

We consider a system $\rm{S}$ interacting with a heat bath $\rm{B}$ at inverse temperature $\beta$, described by the Hamiltonian
\beq
H_{\text{tot}}(t)=H_{\rm{S}}(t)+H_{\rm{B}}+V_{\rm{SB}}(t).
\eeq
The system is assumed to be externally driven out of equilibrium with work being performed, as schematically illustrated in Fig.~\ref{fig:setup}. We note that in our setup the external drive is represented by classical parameters through the time dependence of the Hamiltonian, while there is an alternative formulation that includes the driving system as a part of the quantum system~\cite{Horodecki,Malabarba} (see also Sec.~V~2 of Ref.~\cite{QbookCH25} for a potential problem of this formulation).   

The initial state of the system and the bath is given by the product state
\beq
\rho_{\rm{SB}}(0)=\rho_{\rm{S}}(0)\otimes\rho_{\rm{B}}^{\rm{G}}, \label{setupinitial}
\eeq
where $\rho_{\rm{B}}^{\rm{G}}=e^{-\beta H_{\rm{B}}}/\Tr[e^{-\beta H_{\rm{B}}}]$ is assumed to be the Gibbs distribution of the bath. This is a crucial assumption in deriving the second law and the FT, because the Gibbs distribution is a special state that gives the maximum entropy for a given energy. The composite system evolves in time according to the Schr\"{o}dinger equation, and the unitary time-evolution operator is given by $U_{\rm{SB}}=\text{T}\exp(-\frac{i}{\hbar}\int^{\tau}_{0}dt H_{\text{tot}}(t))$, where $\text{T}$ is the time-ordering operator. The final state is then given by $\rho_{\rm{SB}}(\tau)=U_{\rm{SB}}\rho_{\rm{SB}}(0)U_{\rm{SB}}^{\dagger}$. The following argument is applicable to an arbitrary time-dependent control, as long as the time evolution of the composite system $\rm{SB}$ is unitary. We also note that we do not make any assumption on the size of the bath; it is not necessary to take the thermodynamic limit in the following discussions.

\subsubsection{\label{sec:SLandre}Second law and relative entropy}
We next discuss the derivation of the second law, which is shown to be fundamentally related to the nonnegativity of the quantum relative entropy.  First of all, the total entropy production is defined as
\beq
\Sigma:=\Delta S-\beta Q , \label{totent}
\eeq 
which is a key quantity for quantifying irreversibility in nonequilibrium processes. Here,  $
\Delta S:=S(\rho_{\rm{S}}(\tau))-S(\rho_{\rm{S}}(0))$ is the change in the von Neumann entropy $S(\rho):=-\Tr[\rho\ln\rho]$ of the system, and 
\beq
Q:=\Tr[H_{\rm{B}}\rho_{\rm{B}}^{\rm{G}}]-\Tr[H_{\rm{B}}\rho_{\rm{B}}(\tau)]\label{defheata}
\eeq
is the heat absorbed by the system. Since $-\beta Q$ is interpreted as the entropy change in the bath~\cite{Prigogine}, 
Eq.~(\ref{totent}) quantifies the total entropy produced in the composite system $\rm{SB}$ during nonequilibrium dynamics. 

By using the unitary invariance of the von Neumann entropy, we can relate the total entropy production $\Sigma$ to the quantum relative entropy $S(\rho||\sigma):=\Tr[\rho\ln\rho]-\Tr[\rho\ln\sigma]$~\cite{Nielsen} as 
\beq
\Sigma=S(\rho_{\rm{SB}}(\tau)||\rho_{\rm{S}}(\tau)\otimes\rho_{\rm{B}}^{\rm{G}}).
\eeq
The right-hand side is the relative entropy between the final state $\rho_{\rm{SB}}(\tau)$ of the composite system $\rm{SB}$ and a reference state $\rho_{\rm{S}}(\tau)\otimes\rho_{\rm{B}}^{\rm{G}}$ where only the bath state is replaced by a new Gibbs state, which implies that the concept of entropy production is related to the relaxation of the bath. 
The second law can now be obtained as a direct consequence of the nonnegativity of the quantum relative entropy~\cite{Sagawa12review,Esposito}: 
\beq
\Sigma=\Delta S-\beta Q\geq 0. \label{entsl}
\eeq
Here, the equality is achieved if and only if $\rho_{\rm{SB}}(\tau)=\rho_{\rm{S}}(\tau)\otimes\rho_{\rm{B}}^{\rm{G}}$. The second law~(\ref{entsl}) takes the same form as the conventional Clausius inequality, while it is applicable to arbitrary nonequilibrium initial and final states and includes the von Neumann entropy, instead of the Boltzmann entropy that is defined only for equilibrium states.  
 Inequality~(\ref{entsl}) may also be regarded as a generalized Landauer principle as will be discussed in Sec.~\ref{sec:Landauer}. If we have multiple heat baths, $\beta Q$ in~(\ref{entsl}) should be replaced by $\sum_{i}\beta_{i}Q_{i}$, where $\beta_{i}$ is the inverse temperature of the $i$-th bath and $Q_{i}$ is the heat transfer from the $i$-th bath to the system. 

We next define the work performed on the system through the first law of thermodynamics:
\beq
W:=\Delta E-Q, \label{defWork}
\eeq
where $\Delta E:=\Tr[H_{\rm{S}}(t)\rho_{\rm{S}}(\tau)]-\Tr[H_{\rm{S}}(0)\rho_{\rm{S}}(0)]$ is the energy change of the system. Here, we assume that either (i) a weak coupling between the system and the bath or (ii) $V_{\rm{SB}}(0)=V_{\rm{SB}}(\tau)=0$, such that the change in the interaction energy is negligible. In Eq.~(\ref{defWork}), $W$ quantifies the energy that is injected into the composite system through the time-dependent Hamiltonian of the system via an external control. 
We note that in the strong-coupling regime the definition of the work is given by the energy difference of the composite system including the interaction energy~\cite{Campisi09,Jarzynski04}, and that an extension of the (classical) stochastic thermodynamics has been studied in Refs.~\cite{Seifert16,Jarzynski17,Talkner16}.

Now let us relate the entropy production to the work, and derive a bound on the work. For that purpose, we introduce the nonequilibrium free energy $\mathcal{F}_{\rm{S}}(t)$ of the system, which is motivated by the thermodynamic relation $F=E-TS$ in macroscopic thermodynamics~\cite{Deffner,Esposito2}:
\beq
\mathcal{F}_{\rm{S}}(t):=\Tr[H_{\rm{S}}(t)\rho_{\rm{S}}(t)]-\beta^{-1}S(\rho_{\rm{S}}(t)). \label{noneqfree}
\eeq
This reduces to the equilibrium free energy $F^{\mathrm{eq}}_{\rm{S}}(t):=-\beta^{-1}\ln\Tr[e^{-\beta H_{\rm{S}}(t)}]$ if $\rho_{\rm{S}}(t)=\rho^{\rm{G}}_{\rm{S}}(t)$, where $\rho_{\rm{S}}^{\rm{G}}(t):=e^{-\beta (H_{\rm{S}}(t)-F^{\mathrm{eq}}_{\rm{S}}(t))}$ is the Gibbs distribution of the system at time $t$. In general, (\ref{noneqfree}) is bounded from below as 
\beq
\mathcal{F}_{\rm{S}}(t)=\beta^{-1}S(\rho_{\rm{S}}(t)||\rho_{\rm{S}}^{\rm{G}}(t))+ F^{\mathrm{eq}}_{\rm{S}}(t)\geq F^{\mathrm{eq}}_{\rm{S}}(t). \label{noneqfreeb}
\eeq
From~(\ref{noneqfreeb}), we see that $\mathcal{F}_{\rm{S}}(t)$ quantifies the (asymmetric) distance between $\rho_{\rm{S}}(t)$ and $\rho_{\rm{S}}^{\rm{G}}(t)$.  Using the nonequilibrium free-energy difference $\Delta \mathcal{F}_{\rm{S}}:=\mathcal{F}_{\rm{S}}(\tau)-\mathcal{F}_{\rm{S}}(0)$, the total entropy production is related to the work as
\beq
\Sigma=\beta W-\beta\Delta \mathcal{F}_{\rm{S}}\geq 0. \label{wsecondlaw}
\eeq
Therefore, the second law~(\ref{wsecondlaw}) gives a fundamental lower bound on the work for arbitrary initial and final states of the system. 

If the initial distribution is the Gibbs distribution, i.e., $\rho_{\rm{S}}(0)=\rho_{\rm{S}}^{\rm{G}}(0)$, inequality~(\ref{wsecondlaw}) reproduces the conventional second law
\beq
W-\Delta F^{\mathrm{eq}}_{\rm{S}} \geq 0, \label{wsecondlawa}
\eeq
by noting that $F^{\mathrm{eq}}_{\rm{S}}(0)=\mathcal{F}_{\rm{S}}(0)$, $-F^{\mathrm{eq}}_{\rm{S}}(\tau)\geq -\mathcal{F}_{\rm{S}}(\tau)$, and $\Delta F^{\mathrm{eq}}_{\rm{S}}:=F^{\mathrm{eq}}_{\rm{S}}(\tau)-F^{\mathrm{eq}}_{\rm{S}}(0)$.  
We remark that inequality~(\ref{wsecondlaw}) is tighter than~(\ref{wsecondlawa}), because (\ref{wsecondlawa}) is valid for any final state, but (\ref{wsecondlaw}) has an explicit dependence on the final state $\rho_{\rm{S}}(\tau)$.

\subsubsection{\label{sec:Landauer}Landauer principle}

The Landauer principle~\cite{Landauer,Reeb,Rio11,Sagawa17} gives the fundamental lower bound on the heat emission during the erasure of information. The second law~(\ref{entsl}) is regarded as a general form of the Landauer principle, if we rewrite it as
\beq
-  Q \geq -k_{\rm{B}}T\Delta S, \label{Landauer}
\eeq
where the left-hand side represents the heat emission to the bath, and $-\Delta S$ represents the amount of the erased information. As a simple example, we consider a qubit system and a state transformation from the maximally mixed state $\rho_{\rm{S}}(0)=\frac{1}{2}(|0\rangle\langle 0|+|1\rangle\langle 1|)$ to a pure state  $\rho_{\rm{S}}(\tau)=|0\rangle\langle 0|$. Then, the von Neumann entropy of the system changes from $\ln 2$ to $0$, which is interpreted to be the erasure of one bit of information. From~(\ref{Landauer}), we find that at least $\left|Q\right|=k_{\text{B}}T\ln 2$ of heat should be emitted to the bath, which is nothing but the original Ladauer bound~\cite{Landauer}. 

\subsection{\label{sec:FT}Quantum fluctuation theorems}

\begin{figure}[t]
\begin{center}
\includegraphics[width=.45\textwidth]{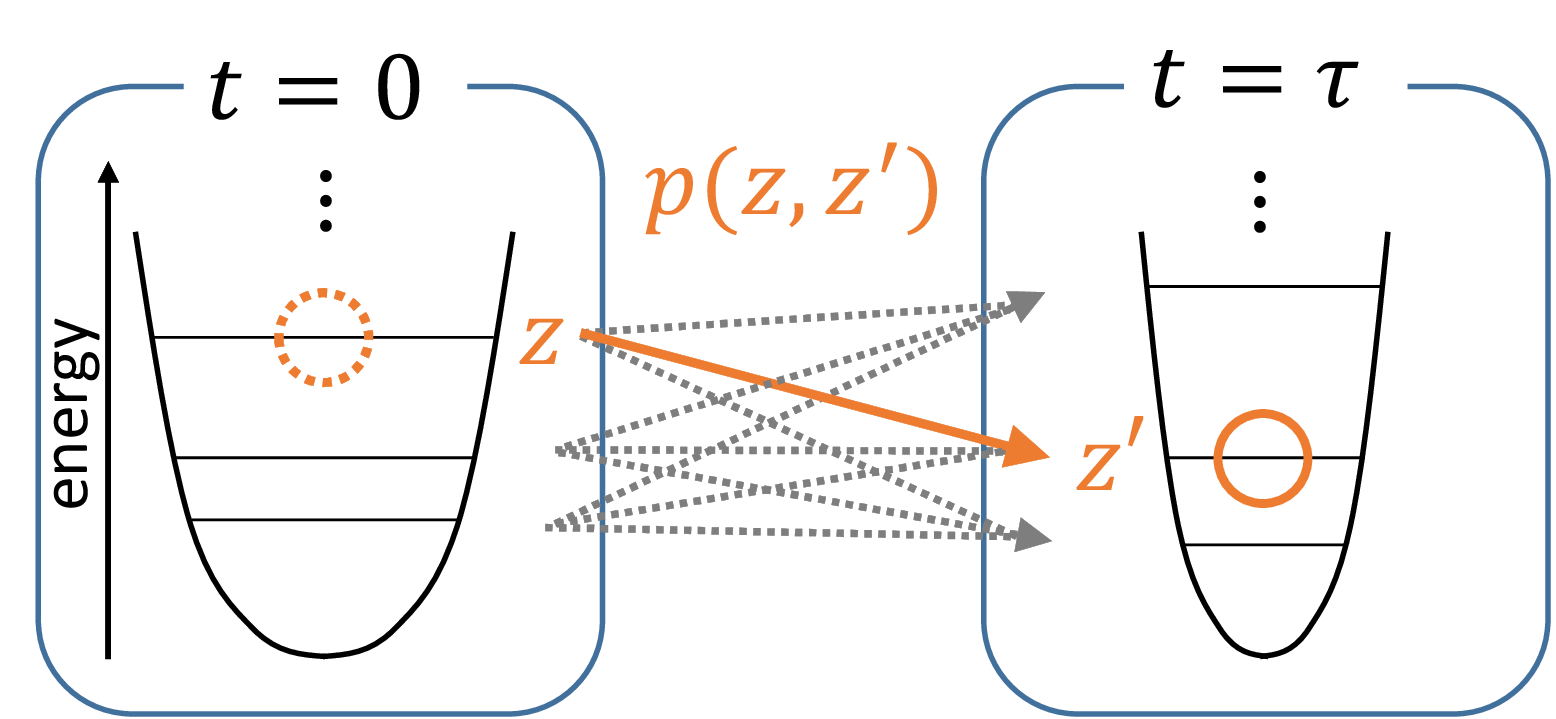}
\caption{ Energy levels of the system in the initial state (left) and the final state (right). The quantum Jarzynski equality (\ref{Jarflucresult}) assumes projective measurements at $t=0$ and $t=\tau$ to specify the initial and final energies. The stochastic entropy production (\ref{totentpro}) is associated with an individual transition $z\rightarrow z'$, where $p(z,z')$ denotes the joint probability distribution. 
}
\label{fig:fluctuation}
\end{center}
\end{figure}

\subsubsection{Stochastic thermodynamic quantities}
The basic setup of quantum fluctuation theorems is the same as that discussed in Sec.~\ref{sec:setup}. 
In addition, we introduce the two-point measurement scheme for the composite system $\rm{SB}$ and define the stochastic entropy production for individual trajectories of transitions (see also Fig.~\ref{fig:fluctuation}). 


The two-point measurement scheme is implemented by two projective measurements at $t=0$ and $t=\tau$ that the bases $\{|\psi_{\rm{S}}(x)\rangle\otimes |E_{\rm{B}}(y)\rangle\}$ and $\{|\psi'_{\rm{S}}(x')\rangle\otimes |E_{\rm{B}}(y')\rangle\}$, respectively. Here, $\{|E_{\rm{B}}(y)\rangle\}$ is the energy eigenbasis of $H_{\rm{B}}$, and $\{|\psi_{\rm{S}}(x)\rangle\}$ is the eigenbasis which diagonalizes the initial density operator of $\rm{S}$:  $\rho_{\rm{S}}(0)=\sum_{x}p_{\rm{S}}(x)|\psi_{\rm{S}}(x)\rangle\langle\psi_{\rm{S}}(x)|$. Similarly, $\{|\psi'_{\rm{S}}(x')\rangle\}$ diagonalizes $\rho_{\rm{S}}(\tau)$ such that $\rho_{\rm{S}}(\tau)=\sum_{x'}p'_{\rm{S}}(x')|\psi'_{\rm{S}}(x')\rangle\langle\psi'_{\rm{S}}(x')|$. Then, the initial measurement gives the stochastic entropy $-\ln p_{\rm{S}}(x)$ of $\rm{S}$ and the energy $E_{\rm{B}}(y)$ of $\rm{B}$. The final measurement gives the same quantities at $t=\tau$.
We then define the stochastic entropy production associated with the transition from $z:=\{x,y\}$ to $z':=\{x',y'\}$ as~\cite{Sagawa12review,Deffner11FT}
\beq
\sigma(z,z'):=\ln p_{\rm{S}}(x)-\ln p'_{\rm{S}}(x')-\beta \left(E_{\rm{B}}(y)-E_{\rm{B}}(y') \right), \label{totentpro}
\eeq
where the first two terms represent the stochastic entropy change of the system and the last two terms give the stochastic heat. 

The joint probability of the measurement outcomes $z$ and $z'$ being observed is given by
\beq
p(z,z')=
p(z'\leftarrow z)p_{\rm{S}}(x)p^{\rm{G}}_{\rm{B}}(y), \label{forwardP}
\eeq
where $p(z'\leftarrow z):=\left| \langle \psi'_{\rm{S}}(x')|\otimes\langle E_{\rm{B}}(y')| U_{\rm{SB}} |\psi_{\rm{S}}(x)\rangle \otimes |E_{\rm{B}}(y)\rangle \right|^{2}$ is the transition probability from $z$ to $z'$, and $p^{\rm{G}}_{\rm{B}}(y)=e^{-\beta E_{\rm{B}}(y)}/Z_{\rm{B}}$ is the Gibbs distribution of the bath. 
By taking the average of $\sigma(z,z')$ with $p(z,z')$, we reproduce the total entropy production defined in~(\ref{totent}): $\Sigma=\sum_{z,z'}p(z,z')\sigma(z,z')$. In addition, the probability density that $\sigma(z,z')$ takes a particular value $\sigma$ is given by
\beq
P(\sigma)=\sum_{z,z'}\delta(\sigma-\sigma(z,z'))p(z,z'), \label{PDent}
\eeq
where $\delta(\cdot)$ is the delta function.



\subsubsection{Quantum fluctuation theorem}

We now discuss the notion of time-reversal symmetry, which directly leads to the quantum FT. We first introduce the backward (time-reversed) protocol and the corresponding backward probability distribution $\tilde{p}(z,z')$ as follows:
\begin{enumerate}
\item We start from a state labeled by $z'=\{x',y'\}$ with a given probability $p_{\rm{S}}'(x')\rho_{\rm{B}}^{\rm{G}}(y')$.
\item The time evolution of the backward protocol is given by $\tilde{U}_{\rm{SB}}$, which connects the initial state of the backward process $|\tilde{\psi}'_{\rm{S}}(x')\rangle\otimes|\tilde{E}_{\rm{B}}(y')\rangle$ to its final state $|\tilde{\psi}_{\rm{S}}(x)\rangle\otimes|\tilde{E}_{\rm{B}}(y)\rangle$. Here, $|\tilde{\phi}\rangle:=\hat{\Theta}|\phi\rangle$ with $\hat{\Theta}$ being the anti-unitary time-reversal operator, and $\tilde{U}_{\rm{SB}}:=\text{T}\exp(-\frac{i}{\hbar}\int_{0}^{\tau}  \tilde{H}_{\text{tot}}(t) dt)$ with $\tilde{H}_{\text{tot}}(t):=\hat{\Theta} H_{\text{tot}}(t)\hat{\Theta}$ being the time-reversed Hamiltonian. 
\end{enumerate}
The backward probability distribution is then given by
\beq
\tilde{p}(z,z')=\tilde{p}(z\leftarrow z')p'_{\rm{S}}(x')\rho_{\rm{B}}^{\rm{G}}(y'), \label{backwardP}
\eeq
where $\tilde{p}(z\leftarrow z'):=|\langle\tilde{\psi}_{\rm{S}}(x)|\otimes\langle\tilde{E}_{\rm{B}}(y)|\tilde{U}_{\rm{SB}}|\tilde{\psi}'_{\rm{S}}(x')\rangle\otimes|\tilde{E}_{\rm{B}}(y')\rangle|^{2}$ 
is the backward transition probability from $z'$ to $z$.

From the unitarity of the time evolution of $\rm{SB}$ and the relation $\hat{\Theta}\tilde{U}_{\rm{SB}}\hat{\Theta}= U^{\dagger}_{\rm{SB}}$, the time-reversal symmetry between the forward and the backward transition probabilities holds: 
\beq
p(z'\leftarrow z)=\tilde{p}(z\leftarrow z'). \label{timereversal}
\eeq
As a consequence, we obtain the detailed FT~\cite{Sagawa12review,Alhambra16}:
\beq
\frac{\tilde{p}(z,z')}{p(z,z')}=e^{-\sigma(z,z')}. \label{detailedFT}
\eeq
We note that essentially the same argument as above has been discussed in Ref.~\cite{Jarzynski00} for classical Liouvillian dynamics.  From Eq.~(\ref{detailedFT}), we derive other types of FTs as follows. 

First, we directly obtain that $P(\sigma)$ in Eq.~(\ref{PDent}) satisfies the Kurchan-Tasaki-Crooks FT:
\beq
\frac{\tilde{P}(-\sigma)}{P(\sigma)}=e^{-\sigma}. \label{CrooksFT}
\eeq
Here, $\tilde{P}(-\sigma):=\sum_{z,z'}\delta\left(\tilde{\sigma}(z,z')+\sigma\right)\tilde{p}(z,z')$ is the probability distribution of entropy production $-\sigma$ in the backward process, where $\tilde{\sigma}(z,z')=\ln \tilde{p}(z,z')-\ln p(z,z')$ is the backward stochastic entropy production. Equality~(\ref{CrooksFT}) shows that the probability of negative entropy production is exponentially small.  


We can further derive the integral quantum FT~\cite{Deffner11FT}:
\beq
\aav{\ee^{-\sigma}}=1. \label{Qfluc}
\eeq
In fact, from Eq.~(\ref{detailedFT}), we have $\aav{\ee^{-\sigma}}=\sum_{z,z'}p(z,z')e^{-\sigma(z,z')}=\sum_{z,z'}\tilde{p}(z,z')=1$, where we use the normalization condition of the backward probability distribution to obtain the last equality. By applying the Jensen inequality $\av{e^{x}}\geq e^{\av{x}}$, Eq.~(\ref{Qfluc}) reproduces the second law~(\ref{entsl}): $\Sigma\geq 0$. By examining the foregoing argument, one can see that the derivation of the second law based on the quantum FT  is essentially the same as that based on the nonnegativity of the quantum relative entropy~\cite{Sagawa12review}.  We also note from Eq.~(\ref{detailedFT}) that $\Sigma$ can be expressed in terms of the classical relative entropy~\cite{Cover} between the forward and backward probabilities: $\Sigma=D(p||\tilde{p}):=\sum_{z,z'}p(z,z')(\ln p(z,z')-\ln \tilde{p}(z,z'))$.

We next express the FT as a property of the characteristic function $\chi(\nu)$ of the entropy production, which is defined as the Fourier transform of $P(\sigma)$:
\beq
\chi(\nu):=\int d\sigma e^{i\nu\sigma}P(\sigma)=\text{Tr}[V_{\nu}^{\dagger}(\tau) U_{\rm{SB}} V_{\nu}(0)(\rho_{\rm{S}}(0)\otimes \rho^{\rm{G}}_{\rm{B}})V_{\nu}(0) U^{\dagger}_{\rm{SB}}V_{\nu}^{\dagger}(\tau) ]. \label{CF}
\eeq
Here, $V_{\nu}(t)=\exp[i\nu(\ln\rho_{\rm{S}}(t)-\beta H_{\rm{B}})/2]$ is a unitary operator that includes the counting field for the full-counting statistics of the entropy production~\cite{Esposito}. Similarly, let us introduce $\tilde{\chi}(\nu)$ as the Fourier transform of $\tilde{P}(\sigma)$. Then, Eq.~(\ref{CrooksFT}) is expressed in terms of the following symmetry of the characteristic function: 
\beq
\chi(\nu)=\tilde{\chi}(-\nu+i).
\eeq
By taking $\nu=i$, we also find that $\chi(i)=\int d\sigma e^{-\sigma}P(\sigma)=\tilde{\chi}(0)=1$, which is nothing but the integral FT~(\ref{Qfluc})~\cite{fluctuation1}. We note that 
the $n$-th cumulant of $\sigma$, written as $\langle \sigma^{n} \rangle_{\rm c}$, can be calculated through the cumulant generating function $K(\nu):=\ln \chi(\nu)$, i.e., $\av{\sigma^{n}}_{\rm c}=(-i\partial_{\nu})^{n}K(\nu)|_{\nu=i}$.

By expanding the cumulant generating function $K(\nu)$ in terms of $\sigma$ up to the second cumulant and by applying Eq.~(\ref{Qfluc}), we obtain the fluctuation-dissipation relation 
\beq
\langle (\sigma-\Sigma)^{2}\rangle=2\Sigma+O(\sigma^{3}), \label{FDT}
\eeq
 where the left-hand side represents the fluctuation of the entropy production and $\Sigma$ on the right-hand side quantifies dissipation. We note that Eq.~(\ref{FDT}) becomes exact when $P(\sigma)$ is the Gaussian distribution. The Onsager reciprocity relation can also be obtained from Eq.~(\ref{detailedFT})~\cite{Saito}. Furthermore, the higher-order extension of these linear relations can systematically be derived from the FT~\cite{Gaspard,Saito2}, and has experimentally been demonstrated in a quantum coherent conductor~\cite{Nakamura}. 



We finally consider a special case in which the initial state of the system is given by the Gibbs distribution $\rho_{\rm{S}}(0)=\rho_{\rm{S}}^{\rm{G}}(0)$ and derive the quantum Jarzynski equality. We first define the stochastic work 
\beq
w(z,z')=\left(E'_{\rm{S}}(x')+E_{\rm{B}}(y')\right)-\left(E_{\rm{S}}(x)+E_{\rm{B}}(y) \right), \label{defwork}
\eeq
where $E_{\rm{S}}(x)$ and $E'_{\rm{S}}(x')$ are the initial and final energies of the system, respectively. 
The work probability distribution is defined as $P(w)=\sum_{z,z'}\delta(w-w(z,z'))p(z,z'),$ where $p(z,z')$ is given in Eq.~(\ref{forwardP}). We note that the stochastic work $w$ cannot be obtained from a projection measurement of a single observable~\cite{Talkner07}. We also note that analytical expressions of the work probability distribution have been obtained for a dragged harmonic oscillator in isolated systems~\cite{Talkner08}, open systems~\cite{Funo17}, and a parametrically driven oscillator in isolated systems~\cite{Deffner10}.

When the initial and final states of the system are given by the Gibbs distributions, the entropy production is given by the difference between the work and the equilibrium free energy: $\sigma(z,z')=\beta(w(z,z')-\Delta F^{\rm{eq}}_{\rm{S}})$.  
The integral FT~(\ref{Qfluc}) then reduces to 
\beq
\aav{\ee^{-\beta(w-\Delta F^{\mathrm{eq}}_{\rm{S}})}}=1, \label{Jarflucresult}
\eeq
which is called the quantum Jarzynski equality~\cite{Tasaki,Kurchan}. We note that Eq.~(\ref{Jarflucresult}) is still valid when the final state of the system deviates from the Gibbs distribution~\cite{Sagawa12review}.


\subsection{\label{sec:c}Gibbs-preserving maps and beyond}

\subsubsection{\label{sec:c:sl}Second law}
In Sec.~\ref{sec:SLandre}, we assumed that the composite system $\rm{SB}$ obeys unitary dynamics. 
In this section, we adopt a slightly different approach, where the bath degrees of freedom are traced out and thermodynamic quantities are defined in terms of the degrees of freedom of the system.

In this situation, we can derive the second law from the monotonicity of the quantum relative entropy~\cite{Lieb,Petz}, which states that  
\beq
S(\rho||\sigma)\geq S(\mathcal{E}(\rho)||\mathcal{E}(\sigma))\label{monotonicity}
\eeq
for any completely-positive and trace-preserving (CPTP) map $\mathcal{E}(\bullet)$. This implies that the CPTP map independently acting on the density operators $\rho$ and $\sigma$ does not increase their distinguishability, and therefore the (asymmetric) distance between $\rho$ and $\sigma$ becomes smaller. 
In what follows, we use~(\ref{monotonicity}) to show the second law  for time-independent Hamiltonians as well as time-dependent ones. \\
 
\noindent {\bf Time-independent control.---} We first suppose that the Hamiltonian of the system is time-independent, i.e., $H_{\rm{S}}(t)=H_{\rm{S}}$. In this case, the time evolution describes a thermal relaxation process. Correspondingly, we assume that the CPTP map on the system is a Gibbs-preserving map whose steady state is the Gibbs distribution, i.e., 
\beq
\mathcal{E}(\rho_{\rm{S}}^{\rm{G}})=\rho_{\rm{S}}^{\rm{G}}. \label{CPTPevolution}
\eeq
The time evolution of the system is given by $\rho'_{\rm{S}}=\mathcal{E}(\rho_{\rm{S}})$, and the 
heat is defined as the increase in the energy of the system:
\beq
Q':=\Tr[H_{\rm{S}}\rho'_{\rm{S}}]-\Tr[H_{\rm{S}}\rho_{\rm{S}}]. \label{defheatb}
\eeq
 From the monotonicity of the relative entropy~(\ref{monotonicity}), we have $S(\rho_{\rm{S}}||\rho_{\rm{S}}^{\rm{G}})\geq S(\mathcal{E}(\rho_{\rm{S}})||\mathcal{E}(\rho_{\rm{S}}^{\rm{G}}))=S(\rho'_{\rm{S}}||\rho_{\rm{S}}^{\rm{G}}),$
 which gives the second law~\cite{Spohn78}
\beq
\Sigma'=\Delta S-\beta Q'\geq 0. \label{SLb}
\eeq 
In general, $Q'$ in Eq.~(\ref{defheatb}) is different from $Q$ in Eq.~(\ref{defheata}) because of the interaction energy, and thus the second law~(\ref{SLb}) is different from~(\ref{entsl}). However, if $\mathcal{E}$ is given by the form of $\mathcal{E}(\rho)=\text{Tr}_{\rm{B}}[ U_{\rm{SB}}(\rho_{\rm{S}}\otimes\rho_{\rm{B}}^{\rm{G}})U^{\dagger}_{\rm{SB}}]$ 
with a special condition
\beq
[H_{\rm{S}}+H_{\rm{B}},U_{\rm{SB}}]=0, \label{Thermalcond}
\eeq
we can show that $Q=Q'$ holds and thus the second laws~(\ref{entsl}) and (\ref{SLb}) become equivalent to each other. 
The condition~(\ref{Thermalcond}) means that the sum of the energies of the system and the bath without the interaction energy is preserved under $\mathcal{E}$, and thus any transition in the system is accompanied by a transition in the bath, where their energy changes have the  same absolute value. In this case, $\mathcal{E}$ is called a thermal operation, which is an extensively used concept in the thermodynamic resource theory~\cite{Horodecki,Fernando2}.  A thermal operation is always a Gibbs-preserving map, but the converse is not necessarily true~\cite{Faist}. \\ 

\noindent {\bf Time-dependent control.---} 
We next consider the case in which the Hamiltonian of the system is time-dependent. We here assume that the dynamics of the system is described by the Markov quantum master equation, which means that we should take the weak-coupling limit between the system and the bath~\cite{Breuer}. The CPTP map $\mathcal{E}$ describes the solution of the master equation from $t=0$ to $\tau$, i.e., $\rho_{\rm{S}}(\tau)=\mathcal{E}(\rho_{\rm{S}}(0))$. A crucial feature of the Markovian dynamics is that we can split $\mathcal{E}$ into the 
product of infinitesimal translations: $\mathcal{E}=\mathcal{E}_{N\Delta t}\circ \cdots \circ\mathcal{E}_{2\Delta t}\circ \mathcal{E}_{\Delta t}$, where $\mathcal{E}_{n\Delta t}$ is a CPTP map and describes an infinitesimal evolution from $t=(n-1)\Delta t$ to $t=n\Delta t$ with $N\Delta t=\tau$ and $\Delta t \to  0$. We assume that the system is driven slowly such that the quantum adiabatic theorem is approximately satisfied~\cite{Albash12} and that $\mathcal{E}_{n\Delta t}$ becomes the Gibbs-preserving map for the instantaneous Hamiltonian of the system at time $t=(n-1)\Delta t$. 

The quantum master equation is given by the Lindblad form
\beq
\partial_{t}\rho_{\rm{S}}(t)=\mathcal{L}_{t}[\rho_{\rm{S}}(t)]=-\frac{i}{\hbar}[H_{\rm{S}}(t),\rho_{\rm{S}}(t)]+\sum_{k,l}\mathcal{D}[L_{kl}(t)]\rho_{\rm{S}}(t), \label{LE}
\eeq
where $\mathcal{D}[L_{kl}(t)]\rho_{\rm{S}}(t)=L_{kl}(t)\rho_{\rm{S}}(t) L_{kl}^{\dagger}(t)-\{L_{kl}^{\dagger}(t)L_{kl}(t),\rho_{\rm{S}}(t)\}/2$ describes dissipation and the Lindblad operator $L_{kl}(t)$ describes a quantum jump from the $k$-th eigenstate to the $l$-th eigenstate of the system: $[L_{kl}(t),H_{\rm{S}}(t)]=\Delta_{kl}(t)L_{kl}(t)$ with $\Delta_{kl}(t)=E^{S}_{t}(k)-E^{S}_{t}(l)$.  We further assume the detailed balance condition $L^{\dagger}_{lk}(t)=L_{kl}(t)e^{-\beta\Delta_{kl}(t)/2}$, which is a sufficient condition to make $\mathcal{E}_{t}$ Gibbs-preserving, 
i.e., $\mathcal{E}_{t}(\rho^{\rm{G}}_{\rm{S}}(t))=\rho^{\rm{G}}_{\rm{S}}(t)$ or equivalently $\mathcal{L}_{t}[\rho_{\rm{S}}^{\rm{G}}(t)]=0$. We again note that $\mathcal{E}_{t}$ is the solution to Eq.~(\ref{LE}) from $t$ to $t+\Delta t$ such that $\rho_{\rm{S}}(t+\Delta t)=\mathcal{E}_{t}(\rho_{\rm{S}}(t))$. 
 
The energy of the system is given by $E_{\rm{S}}(t):=\text{Tr}[\rho_{\rm{S}}(t)H_{\rm{S}}(t)]$, whose time derivative gives $\partial_{t}E_{\rm{S}}(t)=\text{Tr}[\rho_{\rm{S}}(t)\partial_{t}H_{\rm{S}}(t)]+\text{Tr}[(\partial_{t}\rho_{\rm{S}}(t))H_{\rm{S}}(t)]$. From the first law of thermodynamics, we associate these two terms with the work flux and the heat flux:
\beqa
\dot{W}&:=&\Tr[\rho_{\rm{S}}(t)\partial_{t}H_{\rm{S}}(t)], \label{OpenW} \\
\dot{Q}'&:=&\Tr[(\partial_{t}\rho_{\rm{S}}(t))H_{\rm{S}}(t)]=-\sum_{k,l}\Tr[L_{kl}(t)\rho_{\rm{S}}(t)L^{\dagger}_{kl}(t)]\Delta_{kl}(t). \label{OpenQ}
\eeqa
Here, Eq.~(\ref{OpenW}) describes the energy change of the system via the time-dependent control of the Hamiltonian, and Eq.~(\ref{OpenQ}) describes that induced by the effect of the bath. Also, the entropy flux is given by $\dot{S}=-\Tr[\partial_{t}\rho_{\rm{S}}(t)\ln\rho_{\rm{S}}(t)]$. 

The entropy production rate is then defined as $\dot{\Sigma}':=\dot{S}-\beta \dot{Q}'$. Since $\mathcal{E}_{t}$ is CPTP and Gibbs-preserving, we can use the monotonicity of the relative entropy~(\ref{monotonicity}) to obtain the second law in the following form~\cite{Spohn78,Yukawa}:
\beq
\dot{\Sigma}'=\lim_{\Delta t\rightarrow 0}\frac{S(\rho_{\rm{S}}(t)||\rho_{\rm{S}}^{\rm{G}}(t))-S(\rho_{\rm{S}}(t+\Delta t)||\rho_{\rm{S}}^{\rm{G}}(t))}{\Delta t}\geq 0. \label{SLmonotonicity}
\eeq
This is a generalization of~(\ref{SLb}) for time-dependent driving of the Hamiltonian. We note that $\dot{\Sigma}'$ may take negative values for non-Markovian processes~\cite{Esposito10a}.

\subsubsection{\label{sec:QJM}Quantum jump method}
 In Sec.~\ref{sec:c:sl}, we only considered the ensemble-averaged quantities of the system described by the quantum master equation~(\ref{LE}). We next consider the quantum jump method to define the stochastic version of $\Sigma'$, which leads to the quantum FT for open Markovian dynamics.


We start by unraveling the Lindblad master equation~(\ref{LE}) to individual quantum trajectories by the following stochastic Schr\"{o}dinger equation~\cite{Breuer}:
\beq
d|\psi_{t}\rangle=\Bigl[ -\frac{i}{\hbar}H^{\text{eff}}_{\rm{S}}(t)+\frac{1}{2}\sum_{k,l}||L_{kl}(t)|\psi_{t}\rangle||^{2}\Bigr] |\psi_{t}\rangle dt +\sum_{k,l}\Bigl[ \frac{L_{kl}(t)}{||L_{kl}(t)|\psi_{t}\rangle ||}-I\Bigr]|\psi_{t}\rangle dN_{kl}(t). \label{SSE}
\eeq
Here, $dN_{kl}(t)$ is a Poisson increment, which takes on $1$ when the quantum jump described by $L_{kl}(t)$ occurs, and $0$ otherwise. We note that the product of $dN_{kl}(t)$ and other terms is defined by the It\^o form.  Its ensemble average is given by $\mathbb{E}[dN_{kl}(t)]=||L_{kl}(t)|\psi_{t}\rangle||^{2}dt$.  
The second term on the right-hand side of Eq.~(\ref{SSE}) describes such a jump process $|\psi_{t}\rangle \mapsto L_{kl}(t)|\psi_{t}\rangle$ up to  a normalization factor. The no-jump process is described by the first term on the right-hand side, where the system evolves continuously in time via the non-Hermitian effective Hamiltonian $H^{\text{eff}}_{\rm{S}}(t):=H_{\rm{S}}(t)-\frac{i\hbar}{2}\sum_{k,l}L^{\dagger}_{kl}(t)L_{kl}(t).$ Since $[H_{\rm{S}}(t),\sum_{k,l}L^{\dagger}_{kl}(t)L_{kl}(t)]=0$, the non-unitary part of the time evolution generated by $H^{\text{eff}}_{\rm{S}}(t)$ has the effect of reducing the norm of the state vector $|\psi_{t}\rangle$. The density operator of the system is reproduced by taking the ensemble average: $\mathbb{E}[|\psi_{t}\rangle\langle \psi_{t}|]=\rho_{\rm{S}}(t)$. Thus, the ensemble average of Eq.~(\ref{SSE}) reproduces the quantum master equation~(\ref{LE}). 

 We denote the history of a jump process as $\psi^{\tau}_{0}:=\{ (k_{1},l_{1},t_{1}), (k_{2},l_{2},t_{2}), \dots,(k_{N},l_{N},t_{N})\}$ with $0\leq t_{1}\leq t_{2}\leq \dots \leq t_{N}\leq \tau$, where $t_{i}$ is the time at which the $i$-th jump process described by $L_{k_{i}l_{i}}(t_{i})$ occurs. 
From the record of the jump process $\psi^{\tau}_{0}$, we define the 
stochastic heat flux by
\beq
q'[\psi^{\tau}_{0}]:=-\sum_{k,l}\int^{\tau}_{0}dN_{kl}(t)\Delta_{kl}(t),\label{QJheat}
\eeq
which gives $\mathbb{E}[\dot{q}']=\dot{Q}'$ in the ensemble average. As is the case for Sec.~\ref{sec:FT}, let 
$p_{\rm{S}}(x)$ and $p'_{\rm{S}}(x')$ be the diagonal elements of the initial and final density operators, respectively. We then define the stochastic entropy production in a manner similar to Eq.~(\ref{totentpro}):  
\beq
\sigma'[\psi^{\tau}_{0},x',x]:=\ln p_{\rm{S}}(x)-\ln p'_{\rm{S}}(x')-\beta q'[\psi^{\tau}_{0}]. \label{QJent}
\eeq
We note that the forward path probability distribution of the trajectory $(\psi^{\tau}_{0},x',x)$ is given by
\beq
P[\psi^{\tau}_{0},x',x]=\left| \langle \psi'_{\rm{S}}(x')|\mathcal{U}_{\tau,t_{N}}L_{k_{N}l_{N}}(t_{N})\mathcal{U}_{t_{N},t_{N-1}} \dots \mathcal{U}_{t_{2},t_{1}} L_{k_{1}l_{1}}(t_{1}) \mathcal{U}_{t_{1},0} |\psi_{\rm{S}}(x)\rangle \right|^{2}, \label{QJpath}
\eeq
where $\{|\psi_{\rm{S}}(x)\rangle\}$ and $\{|\psi'_{\rm{S}}(x')\rangle\}$ are the eigenbases that diagonalize the initial and final density matrices of the system, respectively, and $\mathcal{U}_{t_{i+1},t_{i}}:=\text{T}\exp[-\frac{i}{\hbar}\int^{t_{i+1}}_{t_{i}} H^{\text{eff}}_{\rm{S}}(t)dt]$ is the non-unitary time evolution generated by the effective Hamiltonian for the no-jump process. 

Using Eqs.~(\ref{QJent}) and (\ref{QJpath}), we can derive the integral FT for $\sigma'$~\cite{Horowitz12,Horowitz13}:
\beq
\langle e^{-\sigma'}\rangle =1. \label{QJFT}
\eeq
It follows from this that the second law~(\ref{SLmonotonicity}) is reproduced by the Jensen inequality. Moreover, by defining the backward probability distribution in the same manner as in Eq.~(\ref{QJpath}), we can show the detailed FT for $\sigma'$~\cite{Horowitz12,Horowitz13}. We note that Eq.~(\ref{QJFT}) has also been derived in Ref.~\cite{Hekking13} for a slightly different setup. 

In the standard derivation of the quantum master equation (\ref{LE}) and its unraveling (\ref{SSE}) from unitary dynamics of the system and the bath, the weak-coupling limit and the rotating wave approximation are assumed~\cite{Breuer}. As a consequence,   Eq.~(\ref{Thermalcond}) is approximately satisfied for any infinitesimal step, and thus each infinitesimal evolution is regarded as an approximate thermal operation. Therefore, the transition in the bath for each step, represented as  $y \to y'$ in Eq.~(\ref{forwardP}), is equivalent to a jump in the system. 
Indeed, the work statistics defined via the two-point measurement scheme for the composite system is found to be equivalent to that defined via the quantum jump method~\cite{Silaev14,Liu14,Liu16}.  

We have assumed that the instantaneous steady state at time $t$ is given by $\rho_{\rm{S}}^{\rm{G}}(t)$. On the other hand, if the steady state is not the Gibbs distribution, $\Sigma$ should be interpreted as a ``nonadiabatic'' entropy production, which satisfies the Hatano-Sasa-type fluctuation theorem~\cite{HatanoSasa,Esposito10}. 
 A quantum analogue of the Hatano-Sasa type relation has been derived and discussed in Refs.~\cite{Horowitz13,Horowitz14,Horowitz15,Horowitz17}. We also note that extensions of Eq.~(\ref{QJFT}) to the case of general CPTP maps have been discussed in Refs.~\cite{Horowitz15,Horowitz17}.

The derivation of Eq.~(\ref{QJFT}) is based on the quantum jump method, which is a quantum counterpart of the Markov jump processes. The quantum Brownian motion described by the Caldeira-Leggett model~\cite{CaldeiraBook} is a quantum counterpart of the Brownian motion described by the Langevin equation. This model is used in Refs.~\cite{Hu,Funo17} to study the quantum FT by the path integral method.

\section{\label{sec:FTF}Fluctuation theorems with measurement and feedback control}
In this section, we generalize the second law and the FT to the case with measurement and feedback control, which is a typical setup of information thermodynamics and can be regarded as a modern formulation of Maxwell's demon.  The demon is the key ingredient of information heat engines, extracting work from the system by utilizing information about fluctuations~\cite{Szilard,Lloyd,Nielsen98,Zurek86,Maxwell,Sagawa12review,Sagawa12reviewa,Parrondo,Maruyama}. The Szilard engine~\cite{Szilard} is the quintessential model of Maxwell's demon, where the feedback controller can extract the work $|W|=k_{\rm{B}}T\ln 2$ from the system by utilizing one bit of information. We note that a quantum extension of the Szilard engine has been studied in Refs.~\cite{Zurek86,qSzilard1,qSzilard3}.




\begin{figure}[t]
\begin{center}
\includegraphics[width=.7\textwidth]{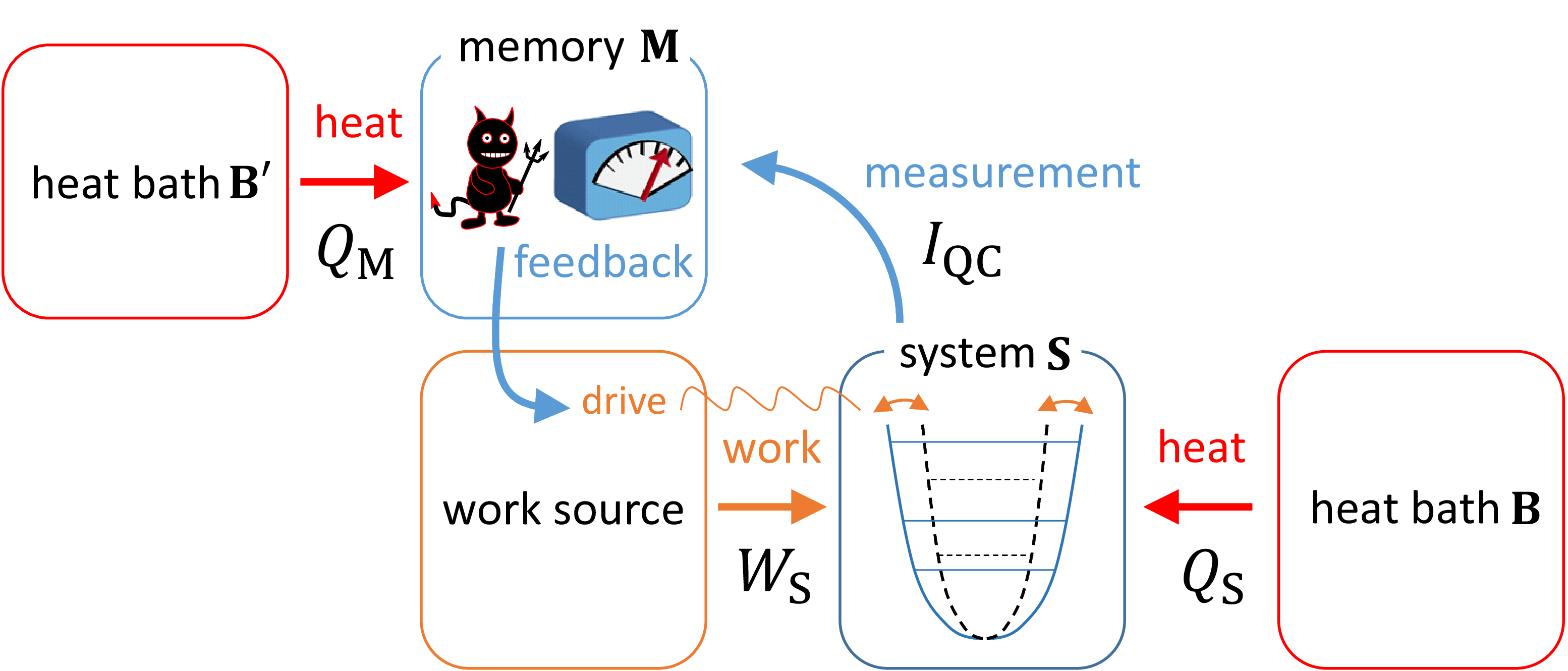}
\caption{ Schematic of the measurement and feedback processes by Maxwell's demon. Depending on the measurement outcome, the demon performs feedback control through the time-dependent control of the system's Hamiltonian. 
}
\label{fig:maxwell}
\end{center}
\end{figure}

\subsection{Setup}
Let us first explain our setup (see also Fig.~\ref{fig:maxwell}), which is an extension of the unitary setup of Sec.~\ref{sec:setup}. We assume that the initial state is given by $\rho_{\rm{S}}\otimes\rho_{\rm{M}}\otimes \rho_{\rm{B}}^{\rm{G}}\otimes\rho_{\rm{B}'}^{\rm{G}}$, where $\rm{S}$ is the controlled system, $\rm{M}$ is the memory that records the measurement outcome, and $\rm{B}$ $(\rm{B}')$ is the heat bath which interacts with $\rm{S}$ ($\rm{M}$). Here, $\rho_{\rm{B}}^{\rm{G}}$ and $\rho_{\rm{B}'}^{\rm{G}}$ are the Gibbs distributions of $\rm{B}$ and $\rm{B}'$ at the same inverse temperature $\beta$. Note that $\rm{M}$ plays the role of Maxwell's demon. For simplicity, we assume that the initial state of M is a pure state: $\rho_{\rm{M}}=|\psi_{\rm{M}}\rangle\langle\psi_{\rm{M}}|$, while the extension to a mixed state is straightforward. Then, the composite system $\rm{SMBB'}$ obeys the following time evolution.

\noindent {\it Measurement process.---}
For simplicity, we assume that $\rm{B}$ is uncoupled to the system during the measurement. A general quantum measurement on $\rm{S}$ is realized by entangling $\rm{S}$ and $\rm{M}$ through a unitary transformation $U_{\rm{SMB'}}$, where the interaction with $\rm{B'}$ is also included. Then, a projection $P_{\rm{M}}(k)=|\varphi_{\rm{M}}(k)\rangle\langle \varphi_{\rm{M}}(k)|$ on $\rm{M}$ follows. The density operator of $\rm{SMB'}$ after the measurement for a given measurement outcome $k$ takes the form 
\beq
\rho'_{\rm{SMB'}}(k)=\frac{1}{p_{k}}P_{\rm{M}}(k)U_{\rm{SMB'}}(\rho_{\rm{S}}\otimes\rho_{\rm{M}}\otimes \rho_{\rm{B'}}^{\rm{G}})U_{\rm{SMB'}}^{\dagger}P_{\rm{M}}(k), 
\eeq
where $p_{k}:=\text{Tr}[P_{\rm{M}}(k)U_{\rm{SMB'}}(\rho_{\rm{S}}\otimes\rho_{\rm{M}}\otimes \rho_{\rm{B'}}^{\rm{G}})U_{\rm{SMB'}}^{\dagger}]$ is the probability of the outcome $k$. 
For simplicity, we assume that the effect of the measurement is described by
\beq
\rho'_{\rm{S}}(k):=\Tr_{\rm{MB'}}[\rho'_{\rm{SMB'}}(k)]=\frac{1}{p_{k}}M_{k}\rho_{\rm{S}}M^{\dagger }_{k}, \label{postmeasstate}
\eeq
where $M_k$ is the Kraus operator satisfying $\sum_k M_k^\dagger M_k = I$.  We note that $p_k = {\rm Tr}[M_k^\dagger M_k \rho_{\rm S}]$, where $\{ M_k^\dagger M_k \}$ is called the POVM~\cite{Nielsen}.


\noindent {\it Feedback control.---} 
We suppose that the unitary time evolution operator $U_{\rm{SB}}(k)$ of $\rm{SB}$ depends on the obtained measurement outcome $k$. 
The density operator after the feedback control is given by $\rho''_{\rm{SB}}(k)=U_{\rm{SB}}(k)(\rho'_{\rm{S}}(k)\otimes\rho_{\rm{B}}^{\rm{G}})U^{\dagger}_{\rm{SB}}(k).$


We now introduce the quantum-classical (QC) mutual information~\cite{Sagawa08} (or the information gain~\cite{Ozawa,Groenewold}) which quantifies the obtained information about the system through the measurement process. The QC-mutual information is defined by 
\beq
I_{\rm{QC}}=S(\rho_{\rm{S}})-\sum_{k}p_{k}S(\rho'_{\rm{S}}(k)), \label{QCmutual}
\eeq
which quantifies how the measurement process reduces the randomness of the system quantified by the von Neumann entropy. 
The QC mutual information satisfies $0 \leq I_{\rm QC} \leq S(p)$, where $S(p) :=-\sum_{k}p_{k}\ln p_{k}$ is the Shannon entropy of the measurement outcome.  The upper bound $I_{\rm QC} = S(p)$ is achieved if every
 Kraus operator is a projection that commutes with $\rho_{\rm S}$, i.e., the measurement is error-free and classical~\cite{Ozawa,Sagawa08}.


We remark on the operational meaning of the QC-mutual information. Suppose that the information about a classical probability variable $m$ is encoded in the density operator as $\rho_{\rm{S}}=\sum_{m}q(m)\sigma_{\rm{S}}(m)$, where $\sigma_{\rm{S}}(m)$'s are not necessarily orthogonal to each other. To extract this information, we perform a measurement described by $\{M_{k}\}$ on $\rho_{\rm{S}}$. The joint probability of $m$ and $k$ is then given by $p(k,m):=q(m)\Tr[M^{\dagger}_{k}M_{k}\sigma_{\rm{S}}(m)]$. In this setup, the following inequality holds~\cite{Buscemi}:
\beq
0\leq I_{\text{CL}}\leq  I_{\rm{QC}} , \label{QCbound}
\eeq
where $I_{\text{CL}}=\sum_{k,m}p(k,m)(\ln p(k,m)-\ln [q(m)p(k)])$ is the classical mutual information between $m$ and $k$. Inequality~(\ref{QCbound}) implies that the QC-mutual information gives an upper bound on the accessible classical information that is encoded in the density operator. 

We note that in general Eq.~(\ref{postmeasstate}) becomes $\rho'_{\rm{S}}(k)=p_{k}^{-1}\sum_{l}M_{k,l}\rho_{\rm{S}}M^{\dagger }_{k,l}$, where $l$ labels the transition of $\rm{B'}$. This describes an inefficient measurement, where the randomness of $\rm{M}$ caused by $\rm{B'}$ is transfered to $\rm{S}$ as a measurement backaction. As a result, Eq.~(\ref{QCmutual}) can take on negative values~\cite{Jacobs,Funo13,Naghiloo18}.




\subsection{Second law of information thermodynamics}


We now discuss several generalizations of the second law of thermodynamics that are applicable to the measurement and feedback processes. We define the entropy production-like quantities $\Sigma_{\rm{S}}$ for the system and $\Sigma_{\rm{M}}$ for the memory by adopting a similar definition as in Eq.~(\ref{totent}). The entropy change in $\rm{SB}$ for a given measurement outcome $k$ is quantified by $\Sigma_{\rm{S}}:=\sum_{k}p_{k}S(\rho''_{\rm{S}}(k))-S(\rho_{\rm{S}})-\beta Q_{\rm{S}}$, and $ Q_{\rm{S}}:=\Tr[H_{\rm{B}}\rho_{\rm{B}}^{\rm{G}}]-\sum_{k}p_{k}\Tr[H_{\rm{B}}\rho''_{\rm{B}}(k)]$ is the heat transfer from $\rm{B}$ to $\rm{S}$. Similarly, the entropy change in $\rm{MB'}$ for the measurement process is quantified by $\Sigma_{\rm{M}}:=S(\rho'_{\rm{M}})-S(\rho_{\rm{M}})-\beta  Q_{\rm{M}}$, where $ Q_{\rm{M}}:=\Tr[H_{\rm{B}'}\rho_{\rm{B}'}^{\rm{G}}]-\Tr[H_{\rm{B}'}\rho'_{\rm{B}'}]$ is the heat transfer from $\rm{B}'$ to $\rm{M}$, and $\rho'_{\rm{SMB'}}:=\sum_{k}p_{k}\rho'_{\rm{SMB'}}(k)$. Note that the definitions of $\Sigma_{\rm{S}}$ and $\Sigma_{\rm{M}}$ are not symmetric, because the roles of $\rm{S}$ and $\rm{M}$ are different in the measurement and feedback processes (see Fig.~\ref{fig:maxwell}). 

A special feature of feedback control  lies in the fact that $\Sigma_{\rm S}$ can become negative up to $-I_{\rm QC}$, which corresponds to the additional work extraction by Maxwell's demon as in the Szilard engine.
In contrast, in the measurement process, $\Sigma_{\rm M}$ is bounded from below by $+ I_{\rm QC}$, and thus cannot reach zero if the memory acquires nonzero information.  
These are represented by the generalized second laws which incorporate the QC-mutual information~\cite{Sagawa08,Sagawa09}
\begin{align}
\Sigma_{\rm{S}} \geq -I_{\rm{QC}}, \label{SLsys} \\ 
\Sigma_{\rm{M}} \geq  +I_{\rm{QC}}. \label{SLmem} 
\end{align}
Here, we notice that $I_{\rm{QC}}$ appears with different signs on the right-hand sides of (\ref{SLsys}) and (\ref{SLmem}). Therefore, if we consider the total entropy production of $\rm{SM}$, the QC-mutual information terms are canceled out:  $\Sigma_{\rm{S}}+\Sigma_{\rm{M}}\geq 0$. This implies that Maxwell's demon is indeed consistent with the conventional second law for the total system. 
From the generalized second laws~(\ref{SLsys}) and (\ref{SLmem}), we find that the combination of $\Sigma_{\rm{S}}+I_{\rm{QC}}$ represents the irreversibility in the feedback control process, and $\Sigma_{\rm{M}}-I_{\rm{QC}}$ represents the irreversibility in the measurement process. In this sense, inequalities~(\ref{SLsys}) and (\ref{SLmem}) give stronger restrictions on the entropy production than the ordinary second law for the total system. 
The equality in~(\ref{SLsys}) is achieved by the classical Szilard engine, where $I_{\rm{QC}}=I_{\rm{CL}}=\ln 2$ and $\Sigma_{\rm{S}}=\beta W_{\rm{S}}=-\ln 2$. A more general protocol to achieve the equality in~(\ref{SLsys}) has been discussed in Ref.~\cite{Jacobs}. 
 
We note that the role of purely quantum correlation (i.e., quantum discord) in the setup of Maxwell's demon has been studied in Refs.~\cite{qSzilard2,Zurek03,Funo13pra}. We also note that there is also another formulation of Maxwell's demon, often referred to as an autonomous demon, which has been studied in both the classical~\cite{Mandal12} and the quantum~\cite{Chapman15} regimes. Such autonomous demons and the measurement-feedback setup has been studied in a unified way~\cite{Strasberg}.


\subsection{Quantum fluctuation theorem}
We next consider the quantum FT for measurement and feedback control processes. The stochastic versions of $\Sigma_{\rm{S}}$ and $\Sigma_{\rm{M}}$ respectively are written as $\sigma_{\rm{S}}$ and $\sigma_{\rm{M}}$, which are defined in a manner similar to that for a non-feedback case~(\ref{totentpro}) (see Refs.~\cite{Funo13,Funo15} for the explicit definitions). We also introduce the stochastic QC-mutual information as~\cite{Funo13}
\beq
i_{\rm{QC}}(x,k,x'):=\ln p'_{\rm{S}}(x'|k) -\ln p_{\rm{S}}(x), \label{fluctuatingQC}
\eeq
where $p_{\rm{S}}(x)$ and $p'_{\rm{S}}(x'|k)$ are the diagonal elements of $\rho_{\rm{S}}$ and $\rho'_{\rm{S}}(k)$, respectively. We can easily show that $\langle i_{\rm{QC}}\rangle =I_{\rm{QC}}$. The right-hand side of~(\ref{fluctuatingQC}) quantifies the stochastic entropy difference between the pre-measurement state and the post-measurement state for a given $k$.  
It is worth comparing Eq.~(\ref{fluctuatingQC}) with the classical stochastic mutual information $i_{\rm{CL}}:=\ln p_{\rm{S}}(x|k)-\ln p_{\rm{S}}(x)$~\cite{Sagawa10,Sagawa12}, where $p_{\rm{S}}(x|k)$ is the initial probability distribution of the system for a given measurement outcome $k$. By comparing $i_{\rm{CL}}$ with $i_{\rm{QC}}$, we find that $i_{\rm{QC}}$ contains the effect of the change in the state of the system from $x$ to $x'$ due to the backaction of the quantum measurement.


In terms of $i_{\rm{QC}}$ in Eq.~(\ref{fluctuatingQC}), the integral FTs for the measurement and feedback processes are shown to be~\cite{Funo13,Funo15}
\begin{align}
\langle e^{-\sigma_{\rm{S}}-i_{\rm{QC}}} \rangle &= 1, \label{FTS} \\
\langle e^{-\sigma_{\rm{M}}+i_{\rm{QC}}}\rangle &= 1. \label{FTM}
\end{align}
The detailed FT has also been derived in Ref.~\cite{Funo15}. Using  the Jensen inequality, Eqs.~(\ref{FTS}) and (\ref{FTM}) reproduce the generalized second laws~(\ref{SLsys}) and (\ref{SLmem}), respectively.   
We note that the entropy production of $\rm{SM}$ also satisfies the FT: $\langle e^{-(\sigma_{\rm{S}}+\sigma_{\rm{M}})} \rangle = 1$. 
Now equalities~(\ref{FTS}) and (\ref{FTM}) include the decomposition of the entropy production of $\rm{SM}$ into $\sigma_{\rm{S}}-i_{\rm{QC}}$ and $\sigma_{\rm{M}}+i_{\rm{QC}}$ at the level of individual trajectories, which is consistent with the decomposition in (\ref{SLsys}) and (\ref{SLmem}) at the level of the ensemble average. 

We note that Eqs.~(\ref{FTS}) and (\ref{FTM}) were first derived in classical systems~\cite{Sagawa10,Sagawa12}.  In Refs.~\cite{Gong,Murashita17}, Eq.~(\ref{FTS}) has been derived on the basis of the quantum jump methods discussed in Sec.~\ref{sec:QJM}. The experimental verification of Eq.~(\ref{FTS}) has been done in Ref.~\cite{Koski14} for a classical system and in Refs.~\cite{Masuyama17,Naghiloo18} for quantum systems. 


\section{\label{sec:experiment}Comparison with experiments}

In this section, we make a brief overview of the experimental studies on quantum thermodynamics, with a special focus on the quantum FT, the Landauer principle, and Maxwell's demon. 

\noindent {\bf Quantum FT.---} 
It has been discussed in Refs.~\cite{Dorner13,Mazzola13} that the work statistics can be measured by adopting a Ramsey-type interferometric scheme. This technique has been utilized in an NMR experiment to obtain the work distribution~\cite{Batalhao14}. A theoretical proposal to extend this technique to the case with feedback control has been discussed in Ref.~\cite{Camati18}. Using a trapped ion system, the two-point measurement scheme has been implemented and the quantum Jarzynski equality for an isolated system has been experimentally verified~\cite{An15}. Using a circuit-QED system, the authors of Ref.~\cite{Naghiloo17} have performed continuous measurements to extract work and heat along quantum trajectories of a qubit.

Many of the systems used in quantum thermodynamic experiments can be regarded as isolated. In such a case, the two-point energy measurement on the system, as performed in Refs.~\cite{An15,Masuyama17}, is essential for obtaining the quantum work distribution. If the effect of the bath cannot be ignored, it is experimentally challenging to verify the quantum FT, since one has to measure the heat exchange between the system and the bath. A possible way to overcome this difficulty is to use single-photon detectors and observe photons emitted from the quantum jump processes, which enables us to measure the stochastic heat as discussed in Sec.~\ref{sec:QJM}.

\noindent {\bf Landauer principle.---} The Landauer principle discussed in Sec.~\ref{sec:Landauer} has been experimentally demonstrated in the classical regime by using a colloidal particle~\cite{Berut,colloidal,John,Berut13,John16,John17}, nanomagnets~\cite{Hong16} and a micro-electromechanical cantilever~\cite{Neri16}.  In the quantum regime, the verification of the Landauer principle has been demonstrated in an NMR experiment through measurements of the heat distribution for elementary quantum logic gates~\cite{Celeri16}.

\noindent {\bf Maxwell's demon.---} 
Experimental implementations of the Maxwell's demon in the classical regime have been achieved with colloidal particles~\cite{Toyabe}, single-electron devices~\cite{Koski14,Chida17,Koski15}, and photonic systems~\cite{Vidrighin16}. 
Maxwell's demon has also been experimentally studied in the quantum regime. Using an NMR system, the authors of Ref.~\cite{Camati16} implemented Maxwell's demon and measured the average entropy production and the information gain. Maxwell's demon based on circuit-QED systems has been experimentally demonstrated in Refs.~\cite{Cottet17,Masuyama17,Naghiloo18}. In Ref.~\cite{Cottet17}, the authors studied the output power with coherent interaction between the demon and the system. In Ref.~\cite{Masuyama17}, a quantum non-demolition (QND) projective measurement technique was utilized to measure the stochastic work and the stochastic QC-mutual information. In Ref.~\cite{Naghiloo18}, a weak continuous measurement was performed to acquire information about the system. Maxwell's demon has also been implemented with the NV-center by combining C-NOT gates~\cite{Wang17}.  A multi-photon optical system has been used for work extraction from entangled bipartite and tripartite states~\cite{Ciampini17}.


\section{\label{sec:summary}Concluding remarks}

In this article, we have discussed some key concepts of the second law of thermodynamics and the FT in the quantum regime. There are a number of subjects that we cannot cover in this article for lack of space. Let us finally make a few remarks about them. 



\noindent {\bf Jarzynski equality for general dynamics.---} 
It is interesting to know to what extent the Jarzynski equality is still valid if the dynamics is not unitary. It has been shown that if the dynamics of the system is described by a unital map (i.e., if the CPTP map does not change the identity operator), the Jarzynski equality is unchanged~\cite{Rastegin,Albash}. This includes a situation where the system is subject to phase decoherence but not to energy dissipation~\cite{Quan17}, and also a situation where an isolated system is continuously monitored by a sequence of projective measurements~\cite{Campisi10}.  

For a general CPTP map that is not necessarily unital, the right-hand side of the Jarzynski equality can deviate from unity~\cite{Rastegin,Albash,Goold15,Kafri}, i.e., 
\beq
\langle e^{-\beta(w-\Delta F^{\rm{eq}}_{\rm{S}})} \rangle=\gamma.\label{CPTPFT}
\eeq
A connection between Eq.~(\ref{CPTPFT}) and the Holevo bound has been discussed in Ref.~\cite{Kafri}. In the context of information thermodynamics, the deviation of $\gamma$ from unity also occurs as a consequence of feedback control~\cite{Sagawa10,Morikuni,Toyabe}.


It is also worth noting that the quantum Jarzysnki equality has been generalized to $\mathcal{PT}$-symmetric non-Hermitian quantum mechanics~\cite{Deffner15}.



\noindent {\bf Quantum-classical correspondence.---} 
It is natural to consider connections between classical and quantum stochastic thermodynamics. Along this line, the quantum-classical correspondence for the work distributions has been shown in isolated~\cite{Jarzynski15,Zhu16} and open~\cite{Funo17} systems.

\noindent {\bf Initial coherence between energy eigenstates.---} The role of quantum coherence in the context of FT has been explored quite recently~\cite{Tajima,Lostaglio18,Aberg18,Llobet17,Solinas,Hofer}. For initial states with coherence between energy eigenstates, an extension of the two-point measurement scheme to define the work distribution is not unique~\cite{Tajima,Llobet17}. If we naively apply the two-point measurement scheme, the initial coherence is destroyed during the first energy measurement. On the other hand, we can utilize the full counting statistics~\cite{Solinas,Hofer} without destroying the initial coherence, although the interference effect may lead to negative values of the work distribution. Also, a variant of the FT that fully includes the effect of coherence has been derived~\cite{Aberg18}.

\noindent {\bf Fluctuation theorems for pure thermal bath.---} In Sec.~\ref{sec:FTW}, we make a crucial assumption that the initial state of the heat bath is given by the Gibbs distribution as in Eq.~(\ref{setupinitial}). 
However, motivated by the recent studies of thermalization in isolated quantum systems, especially the eigenstate thermalization hypothesis~\cite{Rigol}, the authors of Refs.~\cite{Iyoda,Kaneko} have considered a situation in which the initial state of the bath is a pure state and shown that the second law and the FT can still hold at least in a short-time regime.


\bigskip

In summary, the quantum FT is one of the most fundamental relations in nonequilibrium statistical mechanics and applicable to a wide range of dynamics, including quantum information processing. In view of the recent progress in quantum thermodynamics, 
we expect that it will further contribute to the developments of quantum technologies and to the design of microscopic devices with low dissipation that would reach the limit set by the second law of thermodynamics. 

\bigskip

\acknowledgements

The authors thank Y. Masuyama for providing a cartoon of Maxwell's demon in Fig.~\ref{fig:maxwell}. K. F. acknowledges supports from the National Science Foundation of China under Grants No.~11375012 and 11534002, and The Recruitment Program of Global Youth Experts of China. M. U. acknowledges support by a Grant-in-Aid for Scientific Research on Innovative Areas Topological Materials Science (KAKENHI Grant No. JP15H05855). T. S. acknowledges supports from JSPS KAKENHI Grant No. JP16H02211 and No. JP25103003. Part of the research reviewed in this chapter was made possible by the COST MP1209 network ``Thermodynamics in the quantum regime''.

\end{document}